\documentstyle[preprint,aps]{revtex}

\tightenlines

\begin{document}
\title{Localized Magnetoplasmon Modes arising from Broken Translational Symmetry
in Semiconductor Superlattices}

\author{John H. Reina$^{\dagger }$, Juan C. Granada$^{\ddagger }\;$ and Neil F.
Johnson$^{\dagger }$}
\address{$^{\dagger }$Physics Department, Clarendon Laboratory, Oxford University,
Oxford OX1 3PU, U.K.\\
$^{\ddagger }$Departamento de F\'{\i}sica, Universidad del Valle, A.A. 25360,
Santiago de Cali, Colombia}
\date{\today }
\maketitle

\begin{abstract}The electromagnetic propagator associated with the
localized collective magnetoplasmon excitations in a semiconductor
superlattice with broken translational symmetry, is calculated
analytically within linear response theory. We discuss the properties of
these collective excitations in both  {\it radiative} and {\it
non-radiative} regimes of the electromagnetic spectra. We find that low
frequency retarded modes arise when the surface density of carriers $\eta
_{d}\;$at the symmetry breaking layer is lower than the density $\eta
_{_{0}}$ at the remaining layers. Otherwise ($\eta _{d}>\eta _{_{0}}$) a
doublet of localized, high-frequency magnetoplasmon-like modes occurs.
The corresponding dispersion law and power spectrum of these modes are
shown.
\end{abstract}

\newpage

\section{INTRODUCTION}

Much work has been devoted to the study of the properties of low-dimensional
heterostructures \cite{Butcher}. Many of the advances are due to the
emergence of new crystal growth techniques, such as molecular-beam epitaxy
(MBE) and metal organic chemical vapour deposition (MOCVD). These
experimental techniques have made possible the fabrication of layered
materials with sharp, high-quality interfaces and with dimensions comparable
to the electron mean free path and the de Broglie wavelength \cite
{Chang,Joyce}. There are many experimental and theoretical reports
concerning the collective excitations that can propagate in such structures,
including magnons, plasmons, magnetoplasmons and polaritons [4-17]. 
Experimental set-ups for detecting some of these collective oscillatory
modes have been reported recently \cite{Kainth,Moresco}. Most of the
theoretical work in this field is concerned with infinite and semi-infinite
superlattices (for a review see \cite{Albuquerque}). For a periodic array of
quantum wells, a plasmon band arises because of the translational symmetry
and the long-range electromagnetic coupling of the two-dimensional (2D)
electron sheets \cite{Das Sarma,Bloss0,Fetter}. The spectrum of the
collective modes becomes richer when an external static magnetic field is
applied along the superlattice axis. In particular, low-frequency undamped
helicons \cite{Tselis,Babiker,Wendler,Vagner} and waveguide-like modes \cite
{Granada0} can be detected. The case of a semi-infinite semiconductor
superlattice was considered by Giuliani and Quinn \cite
{Giuliani0,Giuliani1,Jain}. They found a polariton-type plasmon mode
(Giuliani-Quinn polariton) which exists only for values of the in-plane
wave-vector greater than a certain critical value. This critical value is
determined by the differences between the dielectric permittivities of the
bulk media and the semi-infinite superlattice. The extension of these works
to the case of applied magnetic fields was discussed by Kushwaha \cite
{Kushwaha}. Recently much attention has been paid to the description of the
local modes related with the removal of translational symmetry, as for a
superlattice with a defect layer \cite{Bloss1,Yibing} or a superlattice with
a quasiperiodic region following Fibonacci \cite{Liu0} or Cantor \cite{Liu1}
sequences. The effects of an applied external magnetic field have also been
considered using Green's function techniques \cite{Jc,Jh0}.

Studies of magnetoplasmons in layered heterostructures have mainly focussed
on the dispersion law of collective excitations, however the power spectrum
of such excitations has not been discussed to our knowledge. In the present
work we calculate the retarded photon Green's tensor $g_{jk}$ for the
magnetoplasmon modes arising in superlattices with broken translational
symmetry. We also discuss the dispersion law (by taking into account
retardation effects and the non-homogeneous propagation of these modes) and
the power spectra of low-frequency collective excitations.

The outline of the paper is as follows: In Section II we describe the system
considered here and obtain analytical expressions for the Green's functions
associated with the localized collective oscillations. The corresponding
discussion, which is centered on the properties of the collective
excitations in both non-radiative and radiative regimes of the
electromagnetic spectra, is given in Section III. Analytical and numerical
treatments are presented for the cases of {\it non-retarded} (Sec. III A)
and {\it homogeneous} (Sec. III B) modes of localized magnetoplasmon
oscillations. The behavior of the {\it retarded non-homogeneous} localized
modes is discussed in Sec. III C. Conclusions are given in Sec. IV.

\section{Formalism}

We consider an infinite one-dimensional array of quantum wells forming a
Type-I superlattice with periodicity $d$. The $z-$direction is taken to lie
along the superlattice axis. We assume that all the quantum wells are
uniformly doped with a surface density of carriers $\eta _{_{0}}$, except
the one centered at $z=0$ which is doped with a surface density $\eta
_{d}^{{}}$. An external static magnetic field $B_{0}$ is applied along the $%
z-$direction. We also assume that the distance $d$ between wells is so large
that we can neglect wave-function overlap between wells. Therefore the
quantum wells are coupled only by the electromagnetic interaction associated
with the dynamics of the electron system.

The dynamic magnetoconductivity tensor corresponding to the $n$'th layer is
given by 
$$
\sigma _{ij}(n)=\sigma _{ij}^{s}+(\sigma _{ij}^{d}-\sigma _{ij}^{s})\delta
_{n{0}}\text{,}\eqno(1)
$$
where $\sigma _{ij}^{s}$ ($\sigma _{ij}^{d}$) are the components of the 2D
dynamic magnetoconductivity tensor associated with the electronic planes
with surface density $\eta _{_{0}}$ ($\eta _{d}$). The photon Green's tensor 
$D_{jk}$ corresponding to the dynamics of the electron system satisfies the
following set of differential equations: 
\[
\left( \frac{\omega ^{2}}{c^{2}}\epsilon \delta _{ij}-\frac{\partial ^{2}}{%
\partial x_{i}\partial x_{j}}+\nabla ^{2}\delta _{ij}\right) D_{jk}(\omega ;%
\mbox{\boldmath $r$}_{\scriptscriptstyle\Vert },\mbox{\boldmath $r$}_{%
\scriptscriptstyle\Vert }^{\prime },z,z^{\prime })=4\pi \hbar \delta
_{ik}\delta (\mbox{\boldmath
$r$}-\mbox{\boldmath $r$}^{\prime })-
\]
$$
-\frac{4\pi i\omega }{c^{2}}\Delta (\eta _{d})\sigma
_{ij}D_{jk}^{(0)}(\omega ;\mbox{\boldmath $r$}_{\scriptscriptstyle\Vert },%
\mbox{\boldmath $r$}_{\scriptscriptstyle\Vert }^{\prime },z^{\prime })\delta
(z)-\frac{4\pi i\omega }{c^{2}}\sum\limits_{n=-\infty }^{\infty }\sigma
_{ij}^{{}}D_{jk}^{(n)}(\omega ;\mbox{\boldmath $r$}_{\scriptscriptstyle\Vert
},\mbox{\boldmath $r$}_{\scriptscriptstyle\Vert }^{\prime },z^{\prime
})\delta (z-nd)\text{,}\eqno(2)
$$
where $\omega $ is the frequency of the allowed collective excitations$,$ $%
\mbox{\boldmath $r$}_{\scriptscriptstyle\Vert }$ and $\mbox{\boldmath $r$}_{%
\scriptscriptstyle\Vert }^{\prime }$ are vectors in the $xy-$plane, $%
\mbox{\boldmath
$r$}=(\mbox{\boldmath $r$}_{\scriptscriptstyle\Vert },z)$ and $\Delta (\eta
_{d})=\frac{\eta _{d}}{\eta _{_{0}}}-1$. The superscript ($n$) on the tensor 
$D_{jk}$ denotes the layer under consideration. Given the homogeneity of the
system along the electron layers, we introduce the 2D Fourier transformation 
$$
D_{jk}(\omega ;\mbox{\boldmath $r$},\mbox{\boldmath$r$}^{\prime })=%
{\displaystyle{1 \over (2\pi )^{2}}}%
\displaystyle\int %
d^{2}k_{\scriptscriptstyle\parallel }d_{jk}(\omega ,\mbox{\boldmath
$k$}_{\scriptscriptstyle\parallel },z,z^{\prime })e^{i\mbox{\boldmath $k$}_{%
\scriptscriptstyle\parallel }\cdot (\mbox{\boldmath
$r$}_{\scriptscriptstyle\parallel }\text{ }-\text{ }\mbox{\boldmath
$r$}_{\scriptscriptstyle\parallel }^{^{\prime }})},\eqno(3)
$$
where $\mbox{\boldmath $k$}_{\scriptscriptstyle\parallel }$ is the in-plane
wave vector with components $\left( k_{x},k_{y}\right) $. The isotropy in
the $xy-$plane allows us to perform a coordinate rotation which aligns the $x
$ axis with the direction of $\mbox{\boldmath$k$}_{\scriptscriptstyle%
\parallel }$. This is achieved by the action of the matrix 
$$
\hat{S}\equiv \hat{S}(\mbox{\boldmath $k$}_{\scriptscriptstyle\parallel })=%
\frac{1}{k_{\scriptscriptstyle\parallel }}\left( 
\begin{array}{ccc}
k_{x} & k_{y} & 0 \\ 
-k_{y} & k_{x} & 0 \\ 
0 & 0 & k_{\scriptscriptstyle\parallel }
\end{array}
\right) \text{.}\eqno(4)
$$
Equation (2) hence takes the form\medskip \bigskip 

$\left( 
\begin{array}{ccc}
\frac{\omega ^{2}}{c^{2}}\epsilon +\frac{d^{2}}{dz^{2}} & 0 & -ik_{%
\scriptscriptstyle\Vert }\frac{d}{dz} \\ 
0 & \frac{d^{2}}{dz^{2}}-\kappa ^{2} & 0 \\ 
-ik_{\scriptscriptstyle\Vert }\frac{d}{dz} & 0 & -\kappa ^{2}
\end{array}
\right) \left( 
\begin{array}{ccc}
g_{xx} & g_{xy} & g_{xz} \\ 
g_{yx} & g_{yy} & g_{yz} \\ 
g_{zx} & g_{zy} & g_{zz}
\end{array}
\right) =4\pi \hbar \delta (z-z^{\prime })\left( 
\begin{array}{ccc}
\;1\; & \;0\; & \;0\; \\ 
\;0\; & \;1\; & \;0\; \\ 
\;0\; & \;0\; & \;1\;
\end{array}
\right) -$%
\begin{eqnarray}
&&\frac{4\pi i\omega }{c^{2}}\Delta (\eta _{d})\left( 
\begin{array}{ccc}
\sigma _{xx} & \sigma _{xy} & 0 \\ 
\sigma _{yx} & \sigma _{yy} & 0 \\ 
0 & 0 & 0
\end{array}
\right) \left( 
\begin{array}{ccc}
g_{xx}^{(0)} & g_{xy}^{(0)} & g_{xz}^{(0)} \\ 
g_{yx}^{(0)} & g_{yy}^{(0)} & g_{yz}^{(0)} \\ 
g_{zx}^{(0)} & g_{zy}^{(0)} & g_{zz}^{(0)}
\end{array}
\right) \delta (z) -  \nonumber \\
&&  \nonumber \\
&&\frac{4\pi i\omega }{c^{2}}\sum\limits_{n=-\infty }^{\infty }\left( 
\begin{array}{ccc}
\sigma _{xx} & \sigma _{xy} & 0 \\ 
\sigma _{yx} & \sigma _{yy} & 0 \\ 
0 & 0 & 0
\end{array}
\right) \left( 
\begin{array}{ccc}
g_{xx}^{(n)} & g_{xy}^{(n)} & g_{xz}^{(n)} \\ 
g_{yx}^{(n)} & g_{yy}^{(n)} & g_{yz}^{(n)} \\ 
g_{zx}^{(n)} & g_{zy}^{(n)} & g_{zz}^{(n)}
\end{array}
\right) \delta (z-nd)\text{ }  \eqnum{5}
\end{eqnarray}
where $g_{jk}=\hat{S}_{jm}d_{mn}\hat{S}_{nk}^{-1}$. In this matrix relation,
we introduce the notation $g_{jk}\equiv g_{jk}(\omega ,\mbox{\boldmath
$k$}_{\scriptscriptstyle\parallel };z,z^{\prime })$,\ $g_{jk}^{(n)}\equiv
g_{jk(n)}=g_{jk}(\omega ,\mbox{\boldmath $k$}_{\scriptscriptstyle\Vert
};z=nd,z^{\prime })$ and $\kappa ^{2}=k_{\scriptscriptstyle\Vert }^{2}-\frac{%
\omega ^{2}}{c^{2}}\epsilon $.

Without loss of generality, we consider the solutions of Eq. (5) in the
region $jd<z<(j+1)d,$ with $j\neq 0$ and $0<z^{\prime }<d.$ The solutions
can be represented as follows: 
$$
g_{ik}^{{}}(z)=\left\{ 
\begin{array}{cc}
g_{ik(n)}^{+}e^{-\kappa z}+g_{ik(n)}^{-}e^{\kappa z}\text{ }, & \text{if }z>0%
\text{ ,} \\ 
\tilde{g}_{ik(n)}^{+}e^{\kappa z}+\tilde{g}_{ik(n)}^{-}e^{-\kappa z}\text{ },
& \text{if }z<0\text{ ,}
\end{array}
\right. \eqno(6) 
$$
satisfying the following boundary conditions at the $n$'th layer: 
$$
g_{ik(n)}^{+}e^{-\kappa d}+g_{ik(n)}^{-}e^{\kappa
d}=g_{ik(n+1)}^{+}+g_{ik(n+1)}^{-}\text{,}\eqno(7) 
$$
\ 
\begin{equation}
-\left( 1+\gamma _{xx}\right) g_{xk(n+1)}^{+}+\left( 1-\gamma _{xx}\right)
g_{xk(n+1)}^{-}+g_{xk(n)}^{+}e^{-\kappa d}-g_{xk(n)}^{-}e^{\kappa d}=\gamma
_{xy}\left( g_{yk(n+1)}^{+}+g_{yk(n+1)}^{+}\right),  \eqnum{8}
\end{equation}
\begin{equation}
-\left( 1+\gamma _{yy}\right) g_{yk(n+1)}^{+}+\left( 1-\gamma _{yy}\right)
g_{yk(n+1)}^{-}+g_{yk(n)}^{+}e^{-\kappa d}-g_{yk(n)}^{-}e^{\kappa d}=\frac{%
4\pi i\omega }{\kappa c^{2}}\sigma _{xy}\left(
g_{xk;n+1}^{+}+g_{xk;n+1}^{+}\right) ,  \eqnum{9}
\end{equation}
with 
\begin{equation}
\gamma _{xx}=%
{\displaystyle{4\pi i\kappa  \over \epsilon \omega }}%
\sigma _{xx}\text{ },\quad \gamma _{xy}=%
{\displaystyle{4\pi i\kappa  \over \epsilon \omega }}%
\sigma _{xy}\text{ },\quad \gamma _{yx}=%
{\displaystyle{4\pi i\omega  \over \kappa c^{2}}}%
\sigma _{yx}\text{ },\quad \gamma _{yy}=-%
{\displaystyle{4\pi i\omega  \over \kappa c^{2}}}%
\sigma _{yy}\;\text{.}  \eqnum{10}
\end{equation}
To solve for the local modes in the superlattice considered here, we assume
a solution 
$$
g_{ik(n)}^{\pm }=g_{ik(0)}^{\pm }\exp (-\alpha |n|d\text{),}\eqno(11) 
$$
which decays exponentially as $n\rightarrow \pm \infty $, i.e. far from the
electron layer with density $\eta _{d}$ $(z=0)$. Equation (11) shows that
the attenuation of the fields along the superlattices axis is characterized
by the parameter $\alpha $. The condition for the existence of non-trivial
solutions of Eq. (5) in the regions considered above, yields 
\begin{equation}
\left( \frac{1}{2}\gamma _{xx}S(\kappa ,\alpha )+1\right) \left( \frac{1}{2}%
\gamma _{yy}S(\kappa ,\alpha )+1\right) =-\frac{\gamma _{xy}\gamma _{yx}}{4}%
S^{2}(\kappa ,\alpha )\text{,}  \eqnum{12}
\end{equation}
where 
\begin{equation}
S(\kappa ,\alpha )=\frac{\sinh (\kappa d)}{\cosh (\kappa d)-\cosh (\alpha d)}%
\text{ }  \eqnum{13}
\end{equation}
is the superlattice structure factor. Equation (12) is expressed in a
general form in terms of the magnetoconductivity tensor $\sigma _{ij}$. We
employ the local approximation for $\sigma _{ij}$: 
\begin{equation}
\sigma _{xx}=\sigma _{yy}=i%
{\displaystyle{e^{2}\eta _{_{0}}\omega  \over m(\omega ^{2}-\omega _{c}^{2})}}%
\text{ },\quad \sigma _{xy}=-\sigma _{yx}=%
{\displaystyle{e^{2}\eta _{_{0}}\omega _{c} \over m(\omega ^{2}-\omega _{c}^{2})}}%
\text{.}  \eqnum{14}
\end{equation}
Here $\omega _{c}=\frac{eB_{0}}{mc}$ is the cyclotron frequency of the
two-dimensional carriers. In this approximation, Eq. (12) takes the form 
\begin{equation}
\left( \frac{\kappa d}{2}S(\kappa ,\alpha )-%
{\displaystyle{\omega ^{2}-\omega _{c}^{2} \over \omega _{p}^{2}}}%
\right) \left( \frac{\omega ^{2}d\epsilon }{2c^{2}\kappa }S(\kappa ,\alpha )+%
{\displaystyle{\omega ^{2}-\omega _{c}^{2} \over \omega _{p}^{2}}}%
\right) =\left( \frac{\omega _{c}d\sqrt{\epsilon }}{2c}S(\kappa ,\alpha
)\right) ^{2}\text{,}  \eqnum{15}
\end{equation}
where {\normalsize $\omega _{p}=$}$\sqrt{\frac{4\pi e^{2}\eta _{_{_{0}}}}{%
m\epsilon d}}$ is the three-dimensional plasmon frequency. Equation (15)
would be equivalent to that obtained in Refs. \cite{Bloss0,Tselis} if $%
i\alpha $ were to be replaced by a quasi wavenumber. When each quantum well
is doped with the same surface density, a magnetoplasma band arises which is
related to the periodicity of the system along the superlattice direction.
However if $\eta _{d}\neq \eta _{_{0}}\;$(i.e. if $\alpha $ is a complex
number), then $|\cosh(\alpha d)|>1$: localized magnetoplasmon modes
associated with the broken translational symmetry appear on the outside of
the bulk magnetoplasmon band.

Following a similar procedure to that described in Ref. \cite{Cott-Mar}, it
is straightforward to show that the diagonal components of the Green's
tensor in the region $-d<z<d,$ with $0<z^{\prime }<d$, have the following
form: 
\begin{eqnarray}
g_{xx}(k_{\scriptscriptstyle\Vert },\omega ;z,z^{\prime }) &=&%
{\displaystyle{\pi c^{2}\hbar \kappa  \over \epsilon \omega ^{2}}}%
f_{xy}(k_{\scriptscriptstyle\Vert },\omega ;z,z^{\prime })\text{,} 
\eqnum{16} \\
g_{yy}(k_{\scriptscriptstyle\Vert },\omega ;z,z^{\prime }) &=&-%
{\displaystyle{\pi \hbar  \over \kappa }}%
f_{yx}(k_{\scriptscriptstyle\Vert },\omega ;z,z^{\prime })\text{,} 
\eqnum{17} \\
g_{zz}(k_{\scriptscriptstyle\Vert },\omega ;z,z^{\prime }) &=&-%
{\displaystyle{ik_{\scriptscriptstyle\Vert } \over \kappa ^{2}}}%
{\displaystyle{d \over dz}}%
g_{xz}(k_{\scriptscriptstyle\Vert },\omega ;z,z^{\prime })-%
{\displaystyle{4\pi \hbar  \over \kappa ^{2}}}%
\delta (z-z^{\prime })\text{,}  \eqnum{18} \\
g_{xz}(k_{\scriptscriptstyle\Vert },\omega ;z,z^{\prime }) &=&%
{\displaystyle{i\pi \hbar c^{2}k_{\scriptscriptstyle\Vert } \over \epsilon \omega ^{2}}}%
\zeta _{xz}\text{.}  \eqnum{19}
\end{eqnarray}
In these expressions we have introduced the shorthand notation: 
\begin{eqnarray}
f_{ij} &\equiv &A(\kappa ,\alpha )\left\{ B(\kappa ,\alpha ,z,z^{\prime
})-C(\kappa ,\alpha ,z)D(\kappa ,\alpha ,z^{\prime })%
{\displaystyle{\left[ \left( \alpha _{ii}^{\prime }+%
{\displaystyle{E(\kappa ,\alpha ,z^{\prime }) \over D(\kappa ,\alpha ,z^{\prime })}}\right) \alpha _{jj}+F(\kappa ,\eta )\right]  \over \alpha _{ii}\alpha _{jj}+%
{\displaystyle{1 \over 2}}F(\kappa ,\eta )}}%
\right\} \text{,}  \nonumber \\
&&  \eqnum{20} \\
\zeta _{xz} &\equiv &A(\kappa ,\alpha )\left\{ G(\kappa ,\alpha ,z,z^{\prime
})-C(\kappa ,\alpha ,z)H(\kappa ,\alpha ,z^{\prime })%
{\displaystyle{\left[ \left( \alpha _{xx}^{\prime }+%
{\displaystyle{J(\kappa ,\alpha ,z^{\prime }) \over H(\kappa ,\alpha ,z^{\prime })}}\right) \widetilde{\alpha }_{yy}+F(\kappa ,\eta )\right]  \over \alpha _{xx}\alpha _{yy}+%
{\displaystyle{1 \over 2}}F(\kappa ,\eta )}}%
\right\} \text{,}  \nonumber \\
&&  \eqnum{21}
\end{eqnarray}
where 
\begin{eqnarray}
A(\kappa ,\alpha ) &=&%
{\displaystyle{1 \over e^{\kappa d}-e^{-\alpha d}}}%
,\quad  \eqnum{22} \\
F(\kappa ,\eta ) &=&%
{\displaystyle{\gamma _{xy}\gamma _{yx} \over 2}}%
\left[ \eta \sinh (\kappa d)\right] ^{2}\text{,}  \eqnum{23} \\
\alpha _{ii}^{\prime }(\kappa ,\alpha ) &=&\cosh (\kappa d)-\chi +\eta
\gamma _{ii}\sinh (\kappa d),  \eqnum{24} \\
\alpha _{ii}(\kappa ,\alpha ) &=&\cosh (\kappa d)-\chi +%
{\displaystyle{1 \over 2}}%
\eta \gamma _{ii}\sinh (\kappa d),  \eqnum{25} \\
\widetilde{\alpha }_{yy}(\kappa ,\alpha ) &=&\cosh (\kappa d)-\chi -%
{\displaystyle{1 \over 2}}%
\eta \gamma _{yy}\sinh (\kappa d)\text{,}  \eqnum{26} \\
C(\kappa ,\alpha ,z) &=&e^{-\alpha d}\sinh (\kappa z)-\sinh \left[ \kappa
\left( z-d\right) \right] ,  \eqnum{27} \\
J(\kappa ,\alpha ,z^{\prime }) &=&2\sinh \left[ \kappa (d-z^{\prime })\right]
-e^{-\kappa z^{\prime }-\alpha d}\text{,}  \eqnum{28} \\
E(\kappa ,\alpha ,z^{\prime }) &=&2\cosh \left[ \kappa (d-z^{\prime })\right]
-e^{-\kappa z^{\prime }-\alpha d},  \eqnum{29} \\
D(\kappa ,\alpha ,z^{\prime }) &=&%
{\displaystyle{2\sinh \left[ \kappa (d-z^{\prime })\right] -e^{-\kappa z^{\prime }-\alpha d} \over \sinh (\kappa d)}}%
\text{,}  \eqnum{30} \\
H(\kappa ,\alpha ,z^{\prime }) &=&%
{\displaystyle{2\cosh \left[ \kappa (d-z^{\prime })\right] -e^{-\kappa z^{\prime }-\alpha d} \over \sinh (\kappa d)}}%
,  \eqnum{31} \\
B(\kappa ,\alpha ,z,z^{\prime }) &=&2\left[ \left( e^{\kappa d}-e^{-\alpha
d}\right) e^{-\kappa \left| z-z^{\prime }\right| }-e^{\kappa z}e^{-\kappa
\left| d-z^{\prime }\right| }\right] \text{,}  \eqnum{32} \\
G(\kappa ,\alpha ,z,z^{\prime }) &=&2\left[ \left( e^{\kappa d}-e^{-\alpha
d}\right) e^{-\kappa \left| z-z^{\prime }\right| }sgn(z-z^{\prime
})-e^{\kappa z}e^{-\kappa \left| d-z^{\prime }\right| }\right] \text{ .} 
\eqnum{33}
\end{eqnarray}
Here $\eta \equiv \frac{\eta _{d}}{\eta _{_{0}}}$ and $\chi =e^{-\alpha d}.$
Notice that the poles of Eqs. (16-18) together with Eq. (15) give the
allowed frequencies of the modes which are localized at the defect layer.
They can be found analytically in the {\it non-retarded} limit and for {\it 
homogeneous oscillations} in the retarded region. These cases, together with
retarded non-homogeneous oscillations, are discussed in the next section.

\section{RESULTS AND DISCUSSION}

\subsection{Non-Retarded Localized Modes{\protect\normalsize \ }}

In the non-retarded region of the spectrum $\omega <<\frac{ck_{_{%
\scriptscriptstyle\parallel }}}{\sqrt{\epsilon }}${\normalsize \ }and%
{\normalsize \ }the components of the Green's tensor have poles at:%
{\normalsize 
$$
\omega ^{2}=\omega _{c}^{2}+%
{\displaystyle{k_{\scriptscriptstyle\parallel }d \over 2}}%
\omega _{p}^{2}\left[ \coth \left( k_{\scriptscriptstyle\parallel }d\right)
+\Delta (\eta )\left\{ \coth ^{2}\left( k_{\scriptscriptstyle\parallel
}d\right) +\eta (\eta -2)\right\} ^{\frac{1}{2}}\right] \text{ ,}\eqno(34) 
$$
$$
e^{\alpha d}=\left( 1-%
{\displaystyle{1 \over \eta }}%
\right) \cosh \left( k_{\scriptscriptstyle\parallel }d\right) +\Delta (\eta
)\left\{ \left( 
{\displaystyle{1 \over \eta }}%
-1\right) ^{2}\cosh ^{2}\left( k_{\scriptscriptstyle\parallel }d\right)
+\left( 
{\displaystyle{2 \over \eta }}%
-1\right) \right\} ^{\frac{1}{2}}\text{.}\eqno(35) 
$$}
Here {\normalsize $\Delta (\eta )\equiv \Theta (\eta -1)-\Theta (1-\eta )$, 
}where{\normalsize \ $\Theta (x)$ }is the Heaviside step function. In order
to have a solution describing the local magnetoplasma mode, we need $X\equiv
\cosh (\alpha d)>1$ or $X<1$. In the non-retarded limit, expression (35)
does not depend on the external magnetic field and a solution of $X$ exists
for all values of the in-plane wave-vector $k_{\scriptscriptstyle\parallel }$%
. This behavior contrasts with that of the Giuliani-Quinn surface polariton 
\cite{Giuliani0} which can not be excited in the long wavelength region of
the spectrum. The above general features are illustrated in Figs. 1 and 2,
where we have plotted the dispersion curves $\omega =\omega (k_{%
\scriptscriptstyle\parallel })$ of the localized magnetoplasma mode for
different values of $\eta $. The regions between the dashed lines correspond
to the allowed frequencies of the magnetoplasmons in the ideal system. It
can be seen that the local frequency increases its value with $k_{%
\scriptscriptstyle\parallel }$. When $k_{\scriptscriptstyle\parallel }d\gg 1$%
, the frequency of the local mode approaches the frequency of the
magnetoplasmon oscillations of a single quantum well with surface carrier
density $\eta _{d}$. This result is reasonable since the coupling between
quantum wells is weak in the $k_{\scriptscriptstyle\parallel }d\gg 1$ limit.
Figure 1 corresponds to values of $\eta <1:$ in this case $X<-1$ and the
local frequencies are lower than those corresponding to the edge of the
Brillouin minizone ($X=-1$). The lowest frequency in this case corresponds
to {\normalsize $\omega _{c}$}. In the strong coupling limit ($k_{%
\scriptscriptstyle\parallel }d\ll 1$), the local mode has a frequency whose
dispersion law can be expressed as{\normalsize 
$$
\omega =\omega _{c}+%
{\displaystyle{\pi \sigma _{_{H}} \over \epsilon d}}%
\eta \left( 1-%
{\displaystyle{\eta  \over 2}}%
\right) \left( k_{\scriptscriptstyle\parallel }d\right) ^{2}\text{,}\eqno%
(36) 
$$
}where{\normalsize 
$$
e^{\alpha d}=%
{\displaystyle{1 \over 2}}%
\left( 
{\displaystyle{1 \over \eta }}%
-1\right) \left( k_{\scriptscriptstyle\parallel }d\right) ^{2}-\left( 
{\displaystyle{2 \over \eta }}%
+1\right) \text{,}\eqno(37) 
$$
}with{\normalsize \ $\sigma _{_{H}}=\frac{ec\eta _{_{0}}}{B_{_{0}}}$}.%
{\normalsize \ }In other words, the behavior of the local low-frequency,
non-retarded modes in the strong coupling limit is governed by the Hall
conductance of the 2D electron gas $\sigma _{_{H}}$. We might conclude from
this analysis that the frequency of this mode is always higher than the
cyclotron frequency of the two-dimensional carriers. We will see however
that accounting for retardation effects can alter this result drastically.

Figure 2 corresponds to $\eta >1$. Then $X>1$ and the local frequencies are
higher than those corresponding to the center of the Brillouin minizone ($%
X=1 $). In the strong coupling limit we have{\normalsize 
$$
\omega ^{2}=\omega _{c}^{2}+\omega _{p}^{2}\left[ 1+\left( 
{\displaystyle{1 \over 3}}%
+%
{\displaystyle{\eta  \over 2}}%
(\eta -2)\right) \left( k_{\scriptscriptstyle\parallel }d\right) ^{2}\right] 
\text{,}\eqno(38) 
$$
}where{\normalsize 
$$
e^{\alpha d}=1+%
{\displaystyle{1 \over 2}}%
\left( 
{\displaystyle{1 \over \eta }}%
-1\right) \left( k_{\scriptscriptstyle\parallel }d\right) ^{2}\text{.}\eqno%
(39) 
$$
}We see that the frequency of this local mode approaches the
three-dimensional value {\normalsize $\sqrt{\omega _{p}^{2}+\omega _{c}^{2}}$%
} as {\normalsize $k_{\scriptscriptstyle\parallel }$} approaches zero.

\subsection{Homogeneous Magnetoplasmon Oscillations}

In the above discussion we have neglected the role of retardation effects.
Now we consider these effects for the $k_{\scriptscriptstyle\parallel }=0$
modes, which corresponds to homogeneous (along the planes) oscillations of
the electron system. In this case

\begin{eqnarray}
\omega &=&\pm \omega _{c}+%
{\displaystyle{\omega _{p}^{2}d \over 2c}}%
\left[ \cot \left( 
{\displaystyle{\omega d\sqrt{\epsilon } \over c}}%
\right) +\Delta (\eta )\left\{ \cot ^{2}\left( 
{\displaystyle{\omega d\sqrt{\epsilon } \over c}}%
\right) -\eta (\eta -2)\right\} ^{\frac{1}{2}}\right] ,  \eqnum{40} \\
e^{\alpha d} &=&\left( 1-%
{\displaystyle{1 \over \eta }}%
\right) \cos \left( 
{\displaystyle{\omega d\sqrt{\epsilon } \over c}}%
\right) +\Delta (\eta )\left\{ \left( 1-%
{\displaystyle{1 \over \eta }}%
\right) ^{2}\cos ^{2}\left( 
{\displaystyle{\omega d\sqrt{\epsilon } \over c}}%
\right) +\left( 
{\displaystyle{2 \over \eta }}%
-1\right) \right\} ^{\frac{1}{2}}\text{.}  \eqnum{41}
\end{eqnarray}
In contrast with previous non-retarded results, the magnitude {\normalsize $%
e^{\alpha d}$} in the present case shows a dependence on the external
magnetic field {\normalsize $B_{0}$}. This follows from the fact that the
modes described by Eq. (41) owe their existence to retardation effects. In
this sense the origin of these local modes differs markedly from the origin
of the non-retarded ones. In the particular case when {\normalsize $\left|
1-\eta \right| \ll 1$} we find localized modes with frequencies close to the
center or the edge of the Brillouin mini-zone ({\normalsize $X=\pm 1$ }%
respectively). We obtain for these frequencies that{\normalsize 
$$
\omega =\omega _{_{0}}-\Delta (\eta )\left( 1-\eta \right) ^{2}%
{\displaystyle{\omega _{p}^{2}d \over 4c\sqrt{\epsilon }}}%
\text{sin}\left( 
{\displaystyle{\omega _{_{0}}d\sqrt{\epsilon } \over c}}%
\right) \text{,}\eqno(42) 
$$
}where {\normalsize $\omega _{_{0}}$ }is the frequency at which the local
mode enters the bulk band of the waveguide-like magnetoplasmon oscillations
which exist in the ideal system \cite{Granada0}. For $\omega _{c}>\frac{c}{d%
\sqrt{\epsilon }}$ the frequency {\normalsize $\omega _{_{0}}$ }satisfies
the inequality {\normalsize $\omega _{_{0}}<$} $\omega _{c}$. This means,
that the local modes have frequencies lower than the electron cyclotron
frequency. The above features are illustrated in Figs. 3 and 4, where we
have plotted the dependences $\omega =\omega (\eta )$ for different values
of the applied external magnetic field.

Figure 3 shows the dependence $\omega =\omega (\eta )$ for the case $\eta
_{d}<\eta _{_{0}}$. When $\eta _{d}$ tends to zero, the local frequency
approaches the cyclotron frequency. This means that the absence of carriers
at the defect layer leads to the appearance of a local mode with a frequency
equal to the cyclotron frequency of the carriers at the remaining layers. We
also see that for a fixed value of the surface carrier density $\eta _{d}$
at the defect layer, the local frequency increases with the applied magnetic
field. For a given value of the external applied magnetic field $B_{0}$,
however, $\omega $ decreases with increasing $\eta _{d}$. In the case $\eta
_{d}>\eta _{_{0}}$ the frequencies corresponding to localized modes appear
only in the high frequency retarded region (i.e. $\omega \gg \omega _{c})$.
The analysis shows that there exist a large number of doublets of local
modes. Figure 4 displays the dependence $\omega =\omega (\eta )$ for the
first doublet. For a given value of $\eta _{d}$ we found that the frequency
of the lowest (highest) component of this doublet decreases (increases) with
applied magnetic field. For this reason the separation between the
components of the doublet increases considerably with magnetic field.

\subsection{General Dispersion Relations and Power Spectra}

We now focuss on the description of the localized modes in the retarded
region $\left( \text{i.e}.\;\omega >\frac{ck_{\scriptscriptstyle\parallel }}{%
\sqrt{\epsilon }}\right) $. Figure 5(a) shows the behavior of the frequency $%
\omega $ (in units of $\omega _{p}$) as a function of in-plane wave number
(in units of $\frac{1}{d}$) for $\eta =0.5$. The dashed region corresponds
to the bulk helicon band. We see that in the forbidden region, a local
helicon mode appears which is localized in the vicinity of the bulk band
edge. With increasing in-plane wave number $k_{\scriptscriptstyle\parallel
}, $ the frequency of the local modes rises until it reaches the electron
cyclotron frequency; here it loses its local-mode character since its
frequency is now indistinguishable from the frequencies describing the layer
dynamics in the ideal system. Also we note that if $\eta \rightarrow 1,$ the
dispersion curves aproach the edge of the bulk helicon band; if on the other
hand $\eta \rightarrow 0,$ it can be seen from the poles of the Green's
functions (Sec. II) that the frequency corresponding to the dynamics of the
defect layer tends to $\omega _{c}\;$for all values of $k_{\scriptscriptstyle%
\parallel }$. In Fig. 5(b) the first doublet of localized modes
corresponding to $\eta =1.5$ is plotted. The dashed area corresponds to the
bulk magnetoplasma band. We see that the lowest mode of the doublet slowly
increases its frequency with $k_{\scriptscriptstyle\parallel }$ until it
reaches the bulk continuum. The other mode of the doublet increases its
frequency more rapidly than the lowest mode, and tends to the dispersion
relation $\omega =\frac{ck_{\scriptscriptstyle\parallel }}{\sqrt{\epsilon }}$
with increasing $k_{\scriptscriptstyle\parallel }$.

The expressions for the diagonal Fourier components of the photon propagator
enable us to obtain the peaks corresponding to power spectra of such modes;
i.e. the imaginary part of the sum over $g_{ii}$ given by Im$\sum g_{ii}$
where $i=x,y,z$. In Fig. 6(a) we show the power spectra of the lowest
frequency modes for different values of $k_{\scriptscriptstyle\parallel },\;$%
in the regime where the density of carriers at the defect layer satisfies $%
\eta _{d}<\eta _{_{0}}$. Only one peak exists which corresponds to
frequencies lower than the cyclotron frequency $\omega _{c}$. With
increasing in-plane wave vector $k_{\scriptscriptstyle\parallel },$ the
frequency at the maxima of such peaks shifts to the right until it reaches $%
\omega _{c}$. These features are also illustrated in Fig. 5(a). In addition,
the frequency corresponding to the peaks in Fig. 6(a) increases when the
density of carriers at the defect layer diminishes.

For the case $\eta _{d}>\eta _{_{0}},$ there are no peaks with frequencies
satisfying $\omega <\omega _{c}$. In this case the peaks corresponding to
localized modes appear in the high frequency retarded region of spectrum.
The analysis shows that there exist a large number of doublets of local
modes. We limit ourselves to discussion of the features corresponding to the
first doublet. In Fig. 6(b) we show the power spectra for different values
of $k_{\scriptscriptstyle\parallel }$. In the homogeneous case the
separation between the components of the doublet is small in comparison with 
$\omega _{p}$ and it increases with increasing in-plane wave vector. These
results agree with those in Fig. 5(b).

\section{Conclusions}

We have used the formalism of linear response theory in order to consider
the localized modes in a Type-I superlattice with broken translational
symmetry, in the presence of an external static magnetic field. We found
that a low frequency retarded mode arises only when the surface carrier
density $\eta _{d}$ is lower than the corresponding density $\eta _{_{0}}$
at the remaining layers. This mode has frequencies greater than those
corresponding to the edge of the helicon band, but is lower than the
cyclotron frequency. In contrast for $\eta _{d}>\eta _{_{0}},$ we found a
doublet of localized high-frequency magnetoplasmon-type modes. These results
can be extended to consider a distribution of defect layers in a Type-I
superlattice. In this case it is necessary to take into account the
techniques developed for the description of localized states in disordered
systems \cite{Ricka}. To our knowledge there are so far no experimental results
on superlattices with a single defect layer, however we believe that the localized
collective excitations described here can indeed be detected with currently
available technology using Raman spectroscopy experiments: similar experiments 
have been reported recently for the case of GaAs/Al$_{x}$Ga$_{1-x}$As double quantum
well structures \cite{Kainth}, where acoustic and optic plasmon modes of
this electron bilayer system were successfully observed using
electronic Raman scattering. We hope that the results reported in this paper
will stimulate further experimental work on this topic.

J.H.R. thanks the financial support of COLCIENCIAS.

\newpage

\begin{center}
{\normalsize {\large {\bf Figure Captions}}}
\end{center}

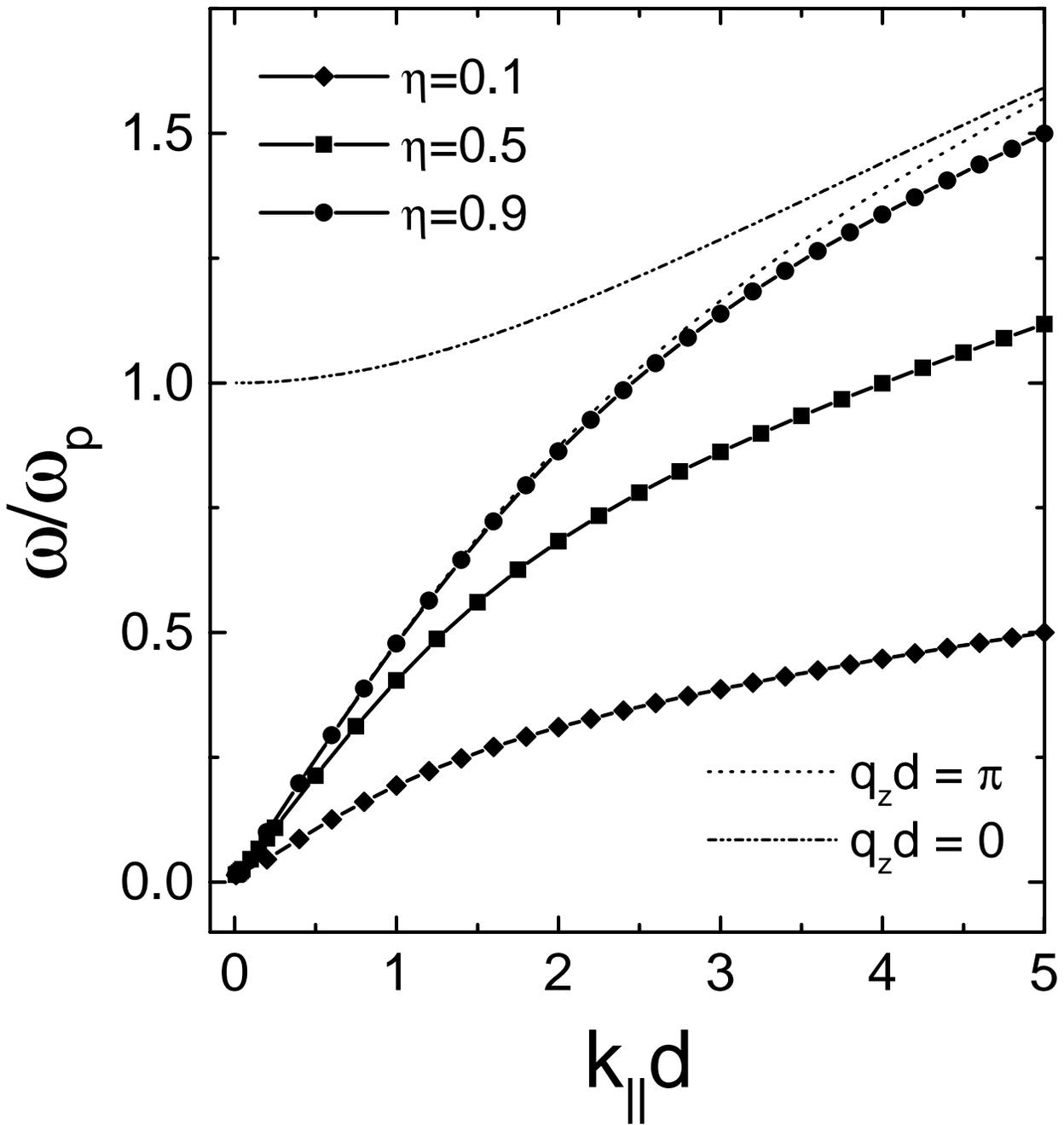
FIGURE 1. Dispersion curves of the local non-retarded magnetoplasmon modes
for different values of the surface carrier density at the defect layer: $%
\eta \equiv \frac{\eta _{d}}{\eta _{_{0}}}=0.1,$ $0.5,$ and $0.9$.$\;$The
regions between the dashed lines correspond to the allowed frequencies of
the magnetoplasmons in the ideal system. $\omega _{c}=0.014\omega _{p}$ and $%
\frac{\omega _{p}d\sqrt{\epsilon }}{c}=2.3.$

FIGURE 2. Same as in Fig. 1 but for $\eta =1.5,$ $\eta =3.0,$ and $5.0$.

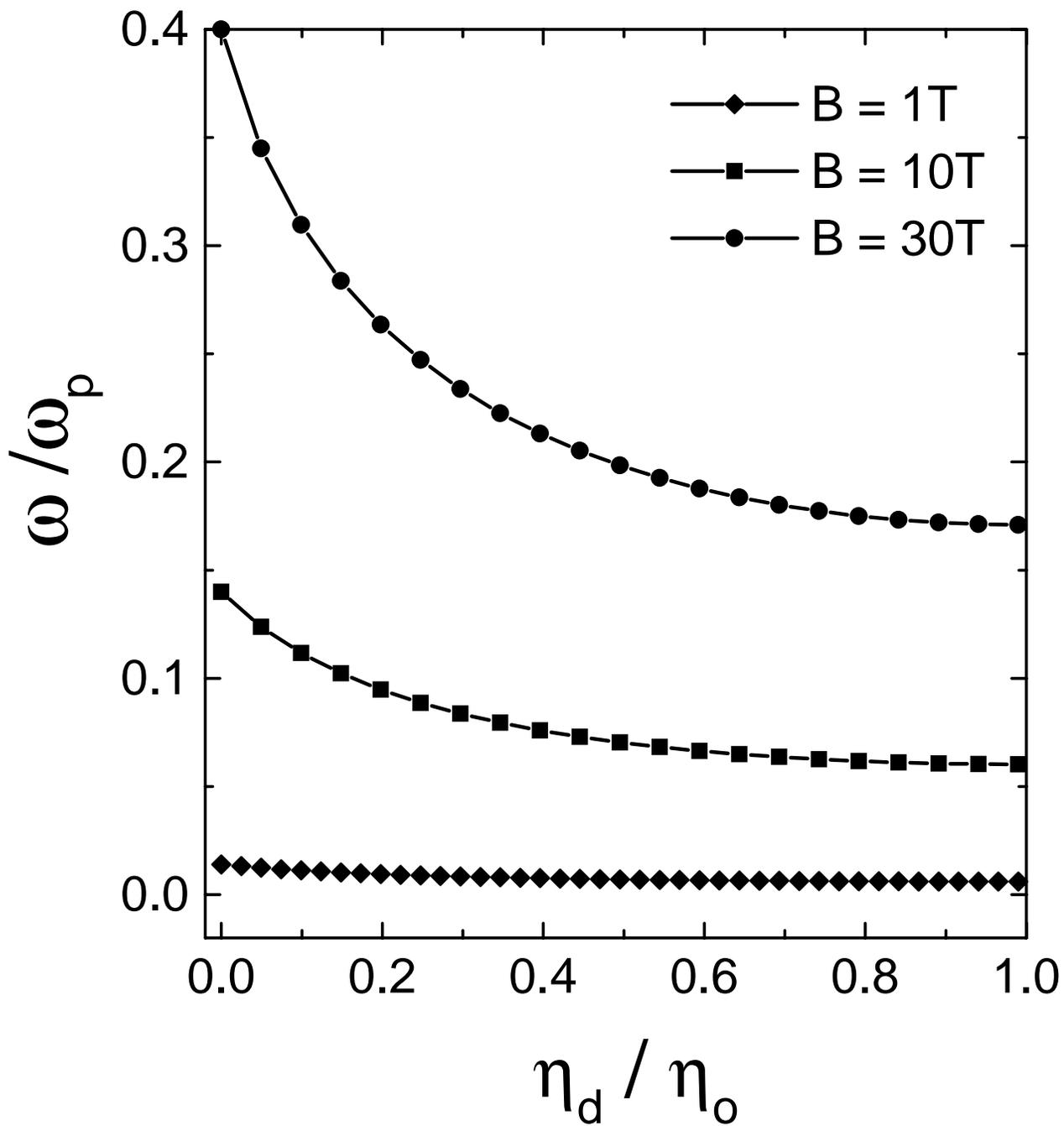
FIGURE 3. Plot of the homogeneous ($k_{\scriptscriptstyle\parallel }=0)$
low-lying local magnetoplasmon frequencies as a function of the surface
carrier density at the defect layer for $\eta _{d}<\eta _{_{0}}$ and for
different values of the applied magnetic field: $\omega _{c}=0.014\omega
_{p}\;(B_{0}=1$T$),$ $0.14\omega _{p}\;(B_{0}=10$T$),$ and $0.41\omega
_{p}\;(B_{0}=30$T$).$ $\frac{\omega _{p}d\sqrt{\epsilon }}{c}=2.3.$

FIGURE 4. Same as in Fig. 3 but for$\;\eta _{d}>\eta _{_{0}}.$

FIGURE 5. (a) Dispersion curves of the localized magnetoplasma modes (solid
lines) for retarded non-homogeneous oscillations $\left( \omega >\frac{ck_{%
\scriptscriptstyle\parallel }}{\sqrt{\epsilon }}\right) .$ $\eta =0.5,$ $%
\omega _{c}=0.014\omega _{p}$ and $\frac{\omega _{p}d\sqrt{\epsilon }}{c}%
=2.3.$ The dashed region corresponds to the bulk helicon band. (b) Same as
in Fig. (a) but for $\eta =1.5.$ The dashed area corresponds to the bulk
magnetoplasmon band.

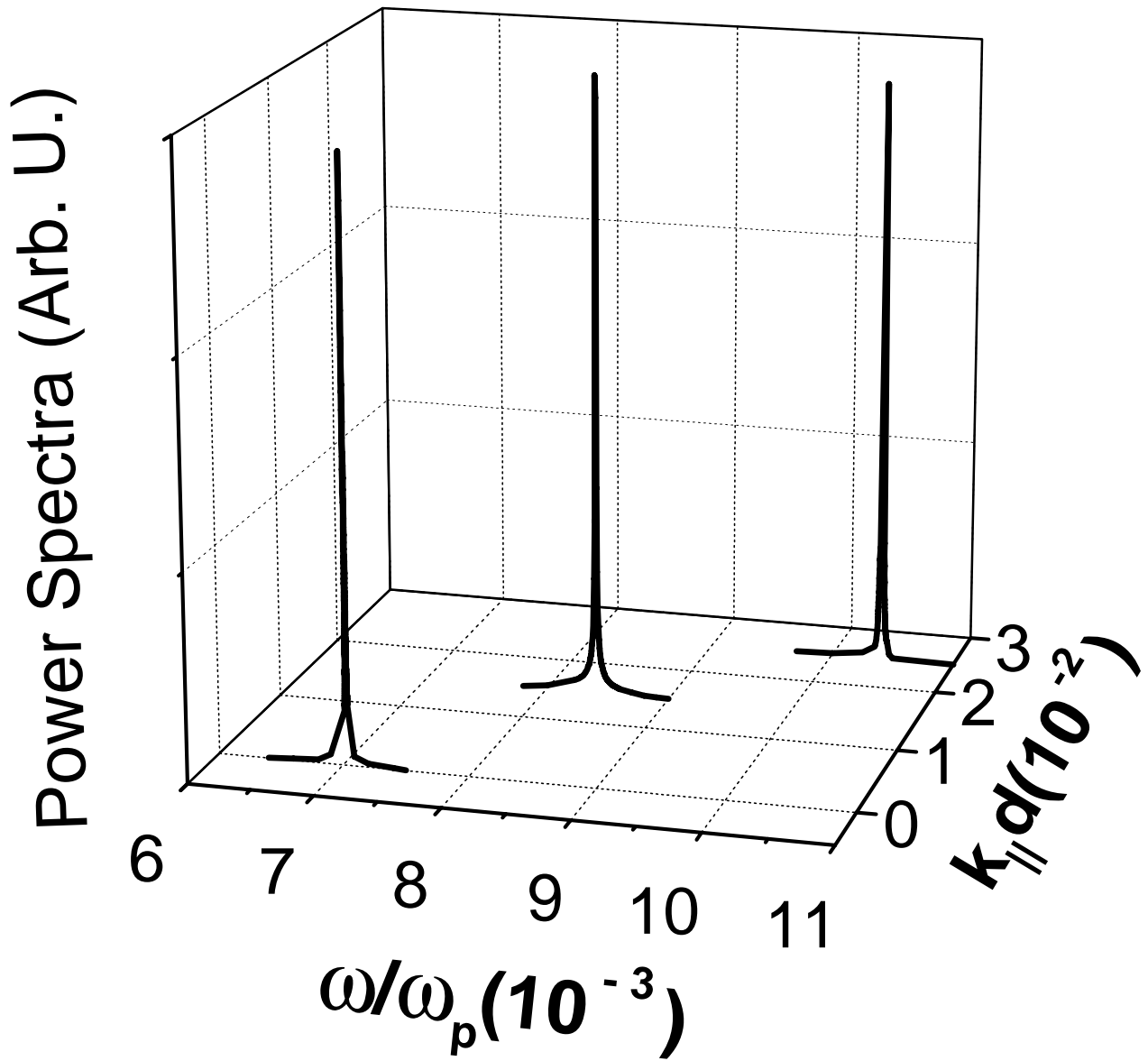
FIGURE 6. (a) Power spectra (in arbitrary units) of low-frequency localized
magnetoplasmon modes as a function of the excitation frequency and the
in-plane wave number in the cases $k_{\scriptscriptstyle\parallel }d=0.0,$ $%
0.015$ and $0.025.$ Input parameters as in Fig. 5(a). (b) Same as in Fig.
(a) but for $k_{\scriptscriptstyle\parallel }d=0.0,$ $1.0$ and $1.5.$ Input
parameters as in Fig. 5(b).


\begin{references}
\bibitem{Butcher}  Stern F 1993 in {\it Physics of Low-dimensional
Semiconductor Structures}, ed P N Butcher, N H March and M P Tosi (Plenum
Press, New York)

\bibitem{Chang}  L L Chang L and K Ploog eds 1985 {\it Molecular Beam
Epitaxy and Heterostructure} (Plenum, New York); E H Parker ed 1985 {\it The
Technology and Physics of Molecular Beam Epitaxy Heterostructure} (Plenum,
New York)

\bibitem{Joyce}  Joyce B A 1985 {\it Rep. Prog. Phys}. {\bf 48} 1637

\bibitem{Bass}  Bass F G and Bulgakov A A 1997 {\it Kinetic and
Electrodynamic Phenomena in Classical and Quantum Semiconductor Superlattices%
} (Nova Science Publishers, New York)

\bibitem{Cottam}  Cottam M G and Tilley D R 1989 {\it Introduction to
Surface and Superlattice Excitations} (Cambridge University Press,
Chichester)

\bibitem{Raj}  Raj R and Tilley D R 1989 {\it The Electrodynamics of
Superlattices}, in The Dielectric Function of Condensed Systems ed R Loudon
and D Kirzhnitz (North Holland, Amsterdam)

\bibitem{Granada}  Granada J C and Oliveira F A 1990 {\it Solid State Commun.%
}{\bf \ 75 }179

\bibitem{Das Sarma}  Das Sarma S and Quinn J J 1982 {\it Phys. Rev. B} {\bf %
25} 7603

\bibitem{Bloss0}  Bloss W L and Brody E M 1982 {\it Solid State Commun.} 
{\bf 43} 523

\bibitem{Tselis}  Tselis A, Gonzalez de la Cruz G and Quinn J J 1983 {\it %
Solid State Commun.} {\bf 47} 43

\bibitem{Giuliani0}  Giuliani G F and Quinn J J 1983 {\it Phys. Rev. Lett.} 
{\bf 51} 919

\bibitem{Giuliani1}  Giuliani G F, Qin G and Quinn J J 1984 {\it Surf. Sci.} 
{\bf 142} 433

\bibitem{Bloss1}  Bloss W L 1991 {\it Phys. Rev. B} {\bf 44} 1105

\bibitem{Jc}  Granada J C 1996 {\it Braz. J. of Physics} {\bf 26} 211

\bibitem{Wendler0}  Wendler L and Grigoryan V G 1999 {\it J. Phys.: Condens.
Matter} {\bf 11} 4199

\bibitem{Kainth}  Kainth D S, Richards D, Bhatti A S, Hughes H P, Simmons M
Y, Linfield E H and Ritchie D A 1999 {\it Phys. Rev. B} {\bf 59} 2095

\bibitem{Moresco}  Moresco F, Rocca M, Hildebrant T and Henzler M 1999 {\it %
Phys. Rev. Lett.} {\bf 83} 2238; Gr\'{e}sillon S {\it et al.} 1999 {\it ibid.%
} {\bf 82} 4520

\bibitem{Albuquerque}  Albuquerque E L and Cottam M G 1993 {\it Phys. Rep.} 
{\bf 240} 383

\bibitem{Fetter}  Fetter A L 1973 {\it Ann. Phys.} {\bf 81} 367

\bibitem{Babiker}  Babiker M 1987 {\it Solid State Commun.} {\bf 64} 983

\bibitem{Wendler}  Wendler L and Kaganov M I 1986 {\it Phys. Status Solidi} 
{\bf K33} 143b

\bibitem{Vagner}  Vagner I D and Bergman D 1987 {\it Phys. Rev. B} {\bf 35}
9856

\bibitem{Granada0}  Granada J C, Kosevich A M and Kosevich Yu A 1990 {\it J.
Phys.: Condens. Matter} {\bf 2} 6279

\bibitem{Jain}  Jain J K and Allen P B 1985 {\it Phys. Rev. B} {\bf 32} 997

\bibitem{Kushwaha}  Kushwaha M 1993 {\it J. Appl. Phys.} {\bf 73} 792

\bibitem{Yibing}  Yibing L and Rocca M 1993 {\it J. Phys.: Condens. Matter} 
{\bf 5} 6597

\bibitem{Liu0}  Liu N H, Feng W and Wu X 1992 {\it J. Phys.: Condens. Matter}
{\bf 4} 9823

\bibitem{Liu1}  Liu N H, Feng W and Wu X 1993 {\it J. Phys.: Condens. Matter}
{\bf 5} 4623

\bibitem{Cott-Mar}  Cottam M G and Maradudin A A 1984 {\it Surface
Excitations} ed V M Agranovich and R Loudon (North-Holland, Amsterdam)

\bibitem{Jh0}  Reina J H and Granada J C 1996 in {\it Surfaces, Vacuum and
their Applications,} Conference Proceedings {\bf 378} 137 (American
Institute of Physics Press); Reina J H 1994 {\it Undergraduate Thesis}
Universidad del Valle (unpublished)

\bibitem{Ricka}  Rickayzen G 1988 {\it Green's Functions and Condensed Matter%
} (Academic Press, London)
\end{references}
\end{document}